# OAM modes characteristics analysis and low-loss transmission based on topological confinement

YIFAN HE,[1] XUCHEN HUA,[1] LEI SHEN,[2] KAI ZHANG,[1] PING WANG,[1] CHENHAO WAN,[1,*] MING TANG[1,*]

[1]*School of Optical and Electronic Information and Wuhan National Laboratory for Optoelectronics, Huazhong University of Science and Technology, Wuhan, Hubei 430074, China*
[2]*State Key Laboratory of Optical Fiber and Cable Manufacture Technology, YOFC, Wuhan, Hubei 430073, China*
* *wanchenhao@hust.edu.cn*
* *tangming@mail.hust.edu.cn*

**Abstract:** The topological confinement is a new mechanism that allows the transmission of cutoff orbital angular momentum (OAM) modes with negligible loss in ring-core fibers (RCFs) and provides a natural immunity against mode coupling. We investigate the influence of fiber design parameters and wavelength on the characteristics of topologically confined modes (TCMs) in step index ring-core fibers (SI-RCFs), and propose a type of graded index ring-core fibers (GI-RCF) with better characteristics. Furthermore, as TCMs occurs in structures with high refractive index difference and are often accompanied by relatively high scattering loss, we fabricate a type of low-loss SI-RCF and observe the stable existence of 24 low-loss TCMs in total. Subsequently, we use an analytical model to estimate the maximum signal-to-noise (SNR) and spectral efficiency (SE) of the fiber, demonstrating its strong capacity advantages.

## 1. Introduction

In the pursuit of increasing the capacity of optical communication, tremendous efforts have been made in the exploration of time, polarization, wavelength/frequency and amplitude/phase multiplexing technologies [1–5]. In the past decade, researchers have attempted to further expand fiber optic communication capabilities by space division multiplexing (SDM) [6,7]. SDM boosts the communication capacity by augmenting the quantity of data paths accessible in a single fiber, encompassing two approaches: core division multiplexing (CDM) and mode division multiplexing (MDM). Despite the significant progress made by CDM [8,9], both the capacity and core count are gradually becoming saturated [6,10]. MDM can theoretically achieve the transmission of a vast number of modes by utilizing orthogonal states with multiple overlapping spaces, and can be categorized into linearly polarized modes [11,12], vector modes [13,14] and OAM modes [15,16] according to the different mode bases. In contrast, OAM modes exhibit better stability and have been extensively investigated [17]. Researchers have put forward numerous multi-mode fiber (MMF) structures that support OAM mode transmission, including air core fibers (ACFs) [18], photonic crystal fibers (PCFs) [19,20], and ring-core fibers (RCFs) [21]. Among these OAM fibers, RCFs are regarded as the structure with highest spatial efficiency. However, two modes with identical effective refractive index $n_{eff}$ undergo mode coupling, which will exert an impact on the stability of mode transmission. This issue can be solved by augmenting the effective refractive index difference $\Delta n_{eff}$ between the two modes [18]. Besides, the mode coupling can also be effectively solved by multiple-input multiple-output digital signal processing (MIMO-DSP) [16,22,23]. Nevertheless, the complexity of mode coupling and MIMO-DSP escalates significantly when more modes and larger differential mode delays are implicated [24].

Topological confinement has been discovered as a new approach to mitigate mode coupling in RCFs. As a new optical guidance mechanism beyond conventional cutoff, topological confinement enables low-loss transmission of multiple OAM cut off modes in SI-RCF without MIMO-DSP, and provides a natural immunity against mode coupling [25]. More specifically, the OAM-induced centrifugal barrier prevents the OAM modes from escaping the fiber and some cutoff high-order modes which are TCMs can violate condition of the total internal reflection (TIR) and behave as bound modes. The losses associated with TCMs are nearly negligible, whereas the losses of high-order modes that may couple with TCMs are several orders of magnitude greater under this mechanism. These high-order modes are unable to survive in the fiber, thereby supporting the transmission of multiple OAM modes without MIMO-DSP. Currently, transmission experiments of TCMs are conducted in a fiber with refractive index difference $\Delta n$ of 0.04 and length of 480m. The average measured loss of TCMs is approximately 5 dB/km. In this paper, we scrutinize the impact of fiber design parameters and wavelength on topological confinement and propose a type of GI-RCF, which shows better characteristics than SI-RCFs. Besides, we fabricate a SI-RCF which enables lower-loss transmission of TCMs and exhibits excellent mode separation capability, and estimate its potential in capacity improvement through an analytical model.

## 2. The influencing factors and mode characteristics analysis of topological confinement

The topological confinement mechanism inherently suppresses the coupling between the first-order radial modes and high-order radial modes. However, for first-order radial modes, a given topological charge $|L|$ can be divided into two pairs of modes depending on different $n_{eff}$: spin-orbit aligned (SOa) modes ($+|L|$, LCP and $-|L|$, RCP) and spin-orbit anti-aligned (SOaa) modes ($-|L|$, LCP and $+|L|$, RCP) [17], and a larger $\Delta n_{eff}$ between them is required to suppress coupling. Besides, fibers are generally designed with a ring-core structure to reduce the number of high-order radial modes. For topological confinement, it is not necessary to restrict the number of high-order radial modes in topological confinement mechanism, which indicates that the inner ring radius $r_1$ of RCFs has no influence on the characteristics of TCMs. Accordingly, the characteristics analysis in terms of fiber structure mainly focuses on two key parameters: the outer ring radius $r_2$ and the refractive index difference $\Delta n$ between the core and cladding. In addition, the wavelength $\lambda$ is taken into account as a parameter. We analyze the effects of above influencing factors on the characteristic of TCMs, mainly including confinement loss, the $\Delta n_{eff}$ between SOa modes and SOaa modes and the effective mode area $A_{eff}$. The characteristics analysis of TCMs in SI-RCFs are simulated in COMSOL Multiphysics.

The effects of different $r_2$ on the characteristics of TCMs in SI-RCFs is shown in Fig. 1. It is concluded that as $r_2$ increases, the topological charge of the first TCM rises gradually and the number of supported TCMs increases steadily. Besides, the increase in $r_2$ leads to a lower $\Delta n_{eff}$ between SOa modes and SOaa modes which is undesirable. In terms of increasing nonlinear resistance, the increase in $r_2$ obviously increases the $A_{eff}$ of the TCMs. The effects of different $\Delta n$ on the characteristics of TCMs in SI-RCFs is shown in Fig. 2. It is concluded that as $\Delta n$ increases, the topological charge of the first TCM rises gradually and the number of supported TCMs increases steadily. Besides, the increase in $\Delta n$ leads to a larger $\Delta n_{eff}$ between SOa modes and SOaa modes. It implies that increasing $\Delta n$ is the optimal choice to avoid mode coupling. And the increase in $\Delta n$ obviously shows a downward trend in the $A_{eff}$ of the TCMs.

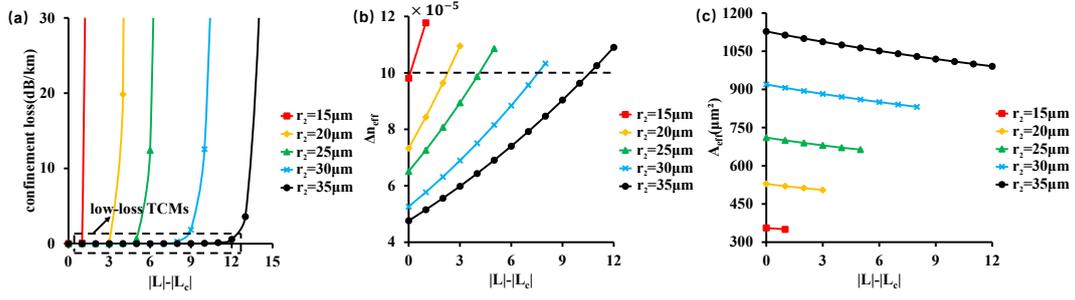

Fig. 1. The influence of $r_2$ on topological confinement and characteristics of the TCMs. (a) The relationship between confinement loss and $|L|-|L_c|$ at $r_2 = 15, 20, 25, 30, 35$ μm with the same $\Delta n = 0.03$ and wavelength of 1550 nm; $L_c$ is the topological charge of the last TIR bound mode ($|L_c| = 14, 19, 25, 30, 35$). (b) The relationship between $\Delta n_{eff}$ and $|L|-|L_c|$ at different $r_2$. (c) The relationship between $A_{eff}$ and $|L|-|L_c|$ at different $r_2$.

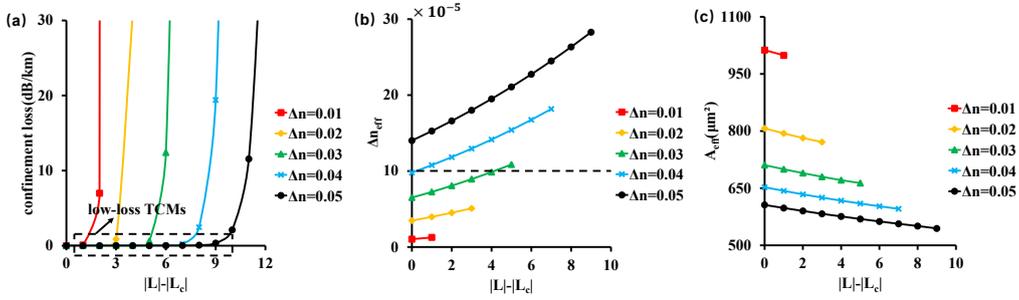

Fig. 2. The influence of $\Delta n$ on topological confinement and characteristics of the TCMs. (a) The relationship between confinement loss and $|L|-|L_c|$ at $\Delta n = 0.01, 0.02, 0.03, 0.04, 0.05$ with the same $r_2 = 25$ μm and wavelength of 1550 nm; $|L_c| = 13, 20, 25, 29, 33$. (b) The relationship between $\Delta n_{eff}$ and $|L|-|L_c|$ at different $\Delta n$. (c) The relationship between $A_{eff}$ and $|L|-|L_c|$ at different $\Delta n$.

Figure 3 shows the number of TCMs under the combined influence of $\Delta n$ and $r_2$. It is observed that topological confinement emerges in SI-RCFs with larger core size (with $r_2$ of at least 20 μm) or higher refractive index difference (with $\Delta n$ of at least 0.02).

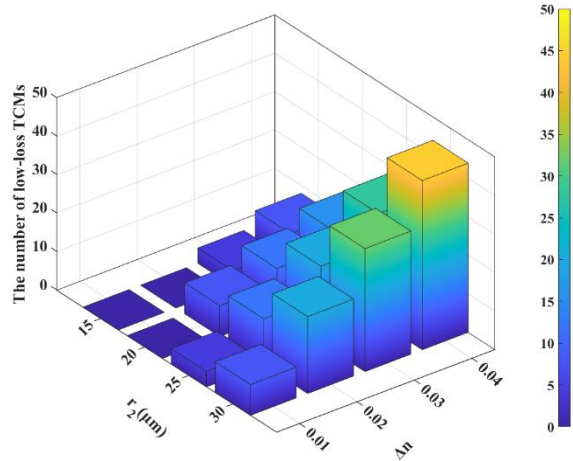

Fig. 3. The number of TCMs under the combined influence of $r_2$ and $\Delta n$. Topological confinement with different $r_2 = 15, 20, 25, 30$ μm and different $\Delta n = 0.01, 0.02, 0.03, 0.04$.

The wavelength $\lambda$ is also a factor to be considered for applying the topological confinement mechanism. Figure 4 shows the impact of different wavelengths on the topological confinement capacity in the same SI-RCF structure. As the wavelength increases, the topological charge of the first TCM and the number of supported TCMs decrease gradually but $\Delta n_{eff}$ and $A_{eff}$ increase steadily which are beneficial for avoiding mode coupling and increasing nonlinear resistance.

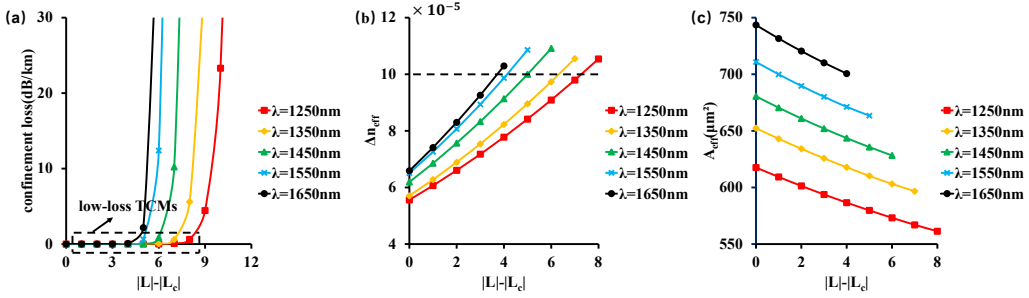

Fig .4. The influence of wavelength on topological confinement and characteristics of the TCMs. (a) The relationship between confinement loss and $|L|-|L_c|$ at $\lambda = 1250, 1350, 1450, 1550, 1650$ nm with the same $r_2 = 25$ μm and $\Delta n = 0.03$ ; $L_c = 32, 29, 27, 25, 23$. (b) The relationship between $\Delta n_{eff}$ and $|L|-|L_c|$ at different wavelengths. (c) The relationship between $A_{eff}$ and $|L|-|L_c|$ at different wavelengths.

The analysis presented above indicates that $A_{eff}$ of TCMs diminishes as the topological charge increases in SI-RCFs, which poses a challenge for increasing nonlinear resistance. In order to solve this issue and validate the universality of topological confinement in MMFs, we explore a new type of graded index ring-core fiber (GI-RCF). The structure formula for the refractive index of the GI-RCF is presented below:

$$n(r) = \begin{cases} n_{co}\left[1-2\Delta\left(\dfrac{r-r_1}{r_2-r_1}\right)^g\right]^{\frac{1}{2}} & (r_1 \leq r \leq r_2) \\ n_{cl} & (0 \leq r < r_1, r > r_2) \end{cases} \quad (1)$$

where $g$ is the index distribution parameter, $\Delta$ is the relative refractive index difference,

$$\Delta = \frac{n_{co}^2 - n_{cl}^2}{2n_{co}^2} \quad (2)$$

The refractive index curves of GI-RCFs with the same structure but different index distribution parameters are shown in Fig. 5a. It is obvious that as $g$ approaches infinity, the fiber becomes a SI-RCF. Figure 5b illustrates the topological confinement capability under different values of $g$. It is concluded that with the increase of $g$, both the topological charge of the first TCM and the number of TCMs gradually increase. Figures 6c and 6d respectively show the variation of $\Delta n_{eff}$ and $A_{eff}$ with topological charge with different $g$. It is shown that when $g = 1$, $\Delta n_{eff}$ between SOa modes and SOaa modes is minimal; however, for $g$ greater than 1, $\Delta n_{eff}$ between SOa modes and SOaa modes becomes nearly identical. In addition, $A_{eff}$ shows an increasing trend at the beginning and then gradually stabilize with the increase of $g$, but $A_{eff}$ starts to decrease when $g>3$. This indicates that SI-RCF represents the most efficient structure concerning the number of modes but exhibits the poorest performance in terms of nonlinearity resistance. In terms of the inhibition of coupling between SOa modes and SOaa modes and the enhancement of nonlinear resistance, the refractive index distribution with $g = 3$ is identified as the optimal choice for GI-RCFs and is demonstrated to have a stable topological confinement capability.

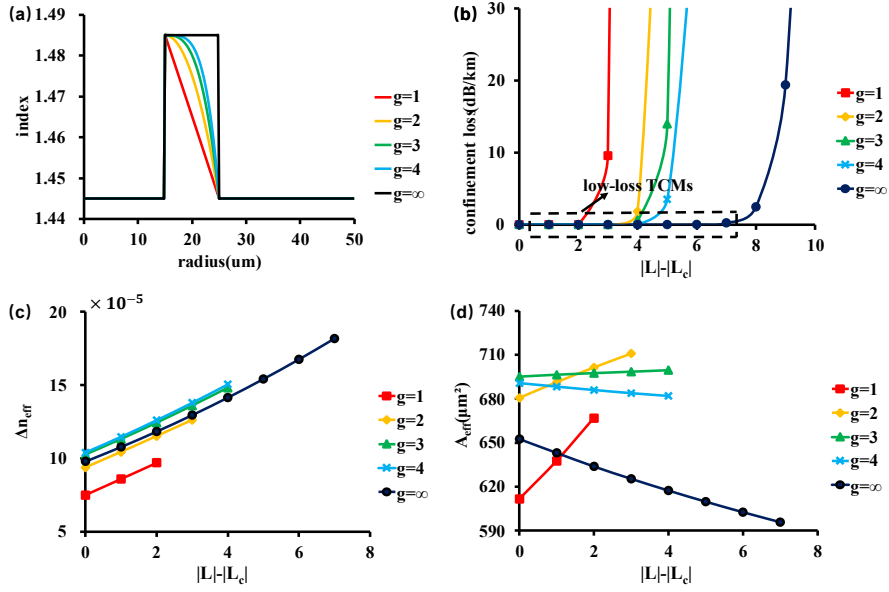

Fig. 5. Topological confinement and characteristics of TCMs in GI-RCFs. (a) The refractive index profile of GI-RCFs at $g=1,2,3,4,\infty$ with the same fiber structure; The inner ring radius $r_1$ is 15 μm, the outer ring radius $r_2$ is 25 μm and the refractive index difference $\Delta n = n_{co} - n_{cl} = 0.04$. (b) The relationship between confinement loss and $|L|-|L_c|$ at different $g$; $L_c = 19, 22, 24, 25, 29$. (c) The relationship between $\Delta n_{eff}$ and $|L|-|L_c|$ at different $g$. (d) The relationship between $A_{eff}$ and $|L|-|L_c|$ at different $g$.

## 3. Transmission experiment of low-loss TCMs

Currently, TCMs face the challenge of high transmission loss when propagating in practical optical fibers, which severely restricts their application prospects in long distance communication. Using a design of low doping in the ring core and negative doping in the cladding with the plasma chemical vapor deposition (PCVD) technique, we fabricate a type of low-loss SI-RCF supporting multiple TCMs. The structure of the low-loss SI-RCF is shown in Fig. 6.

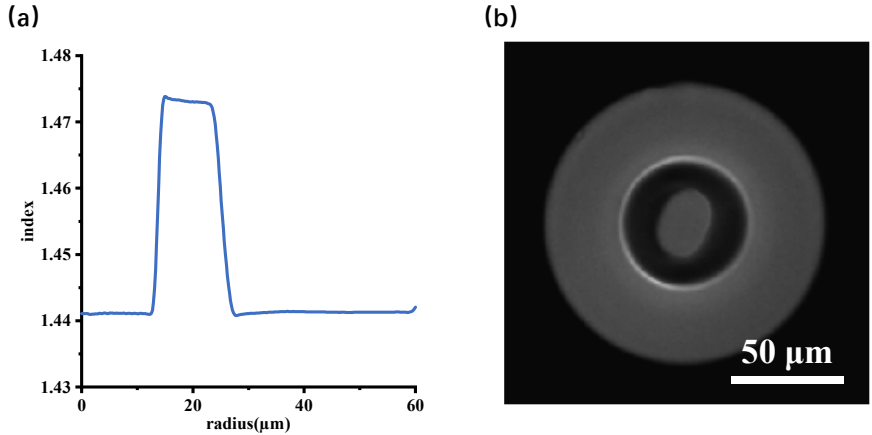

Fig. 6. The structure of the low-loss SI-RCF. (a) The refractive index profile of the fiber. The refractive index difference $\Delta n$ between the core and cladding is about 0.031, where the cladding refractive index is about 1.442 and the core refractive index is 1.473. The outer radius of the core $r_2$ is about 27μm and the inner radius $r_1$ is about 16μm. (b) The optical micrograph of the fiber.

The experimental setup for the transmission of TCMs in the low-loss SI-RCF is shown in Fig. 7. All optical components are arranged sequentially along the optical transmission path. The laser outputs a Gaussian beam with a central wavelength of 1550 nm. The Gaussian beam first passes through a linear polarizer, converting it into a linearly polarized state that can be effectively modulated by the spatial light modulator (SLM). The SLM converts the Gaussian beam into an OAM beam with the target topological charge according to a predefined phase pattern. To eliminate the interference of the unmodulated Gaussian beam caused by pixel gaps in the spatial light modulator, an aperture is subsequently installed to separate the zero-order unmodulated beam from the first-order diffracted OAM beam. Subsequently, a quarter-wave plate converts the linear polarization of the OAM beam into circular polarization suitable for fiber transmission. The beam coupling section consists of two lenses L1 and L2 with different focal lengths and a six-axis translation stage. L1 is a collimating lens with a focal length of 20 cm, used to collimate the diverging OAM beam and reduce the influence of beam divergence on coupling efficiency. L2 is a focusing lens with a focal length of 10 mm, used to precisely couple the collimated OAM beam into the fiber. The fiber input end is mounted on the six-axis translation stage, which enables precise adjustment in lateral, axial, and tilt directions, which is crucial for exciting OAM modes in the fiber. L3 is a collimating lens with a focal length of 20 cm, used to collimate the diverging beam exiting the fiber. The beam finally enters a CCD camera, which captures the far-field pattern of the output beam.

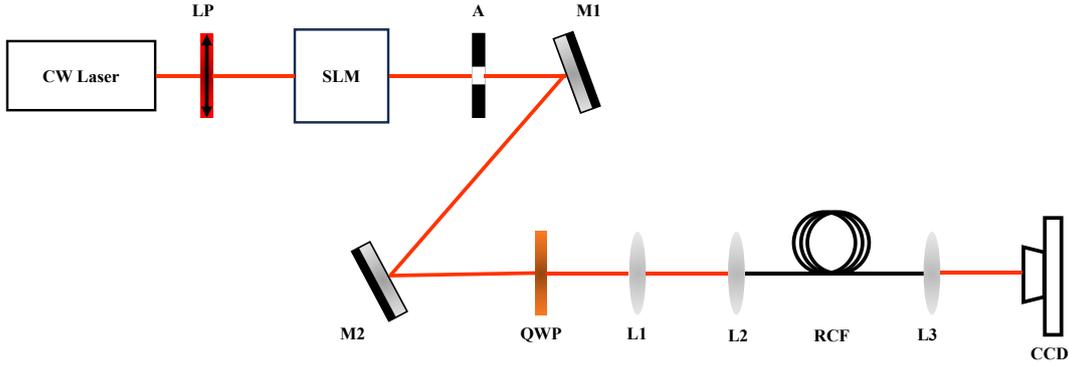

Fig. 7. Experimental setup. CW laser is a 1550nm external cavity laser with continuous wave output. LP: linear polarizer; SLM: spatial light modulator; A: aperture; QWP: quarter wave plate; M: mirror; L1, L2, L3 are lenses with focal lengths of 20cm, 10mm and 20cm, respectively.

We calculate $n_{eff}$ distribution curves of OAM modes versus wavelength the low-loss SI-RCF are shown in Fig. 8. It can be observed that at 1550 nm, the highest bounded OAM mode of highest $|L|$ satisfying the TIR condition in the fiber is 27. When $|L|=28$, the $n_{eff}$ of the corresponding OAM mode becomes lower than the refractive index of the cladding, which no longer satisfies TIR condition and thus forms the first TCM in the fiber. Figure 9 illustrates the mechanism of topological confinement suppressing mode coupling in the low-loss SI-RCF. At 1550nm, the bounded mode with $|L|=27, m=1$ and the high-order mode with $|L|=11, m=4$ exhibit identical $n_{eff}$. Since their confinement losses are on the order of $10^{-11}$ dB/m, mode coupling occurs inevitably, resulting in the distorted coupled pattern shown after fiber transmission in Fig. 9(a). In contrast, for the cutoff TCM with $|L|=33, m=1$ and and the high-order mode with $|L|=25, m=3$ exhibit identical $n_{eff}$. However, since their confinement losses are on the order of $10^{-3}$ and $10^{2}$ dB/m respectively, a loss difference of up to five orders of magnitude is generated. The high-order radial modes decay rapidly during transmission, thus effectively suppressing mode coupling. A clear annular pattern is observed after fiber transmission as shown in Fig. 9(b).

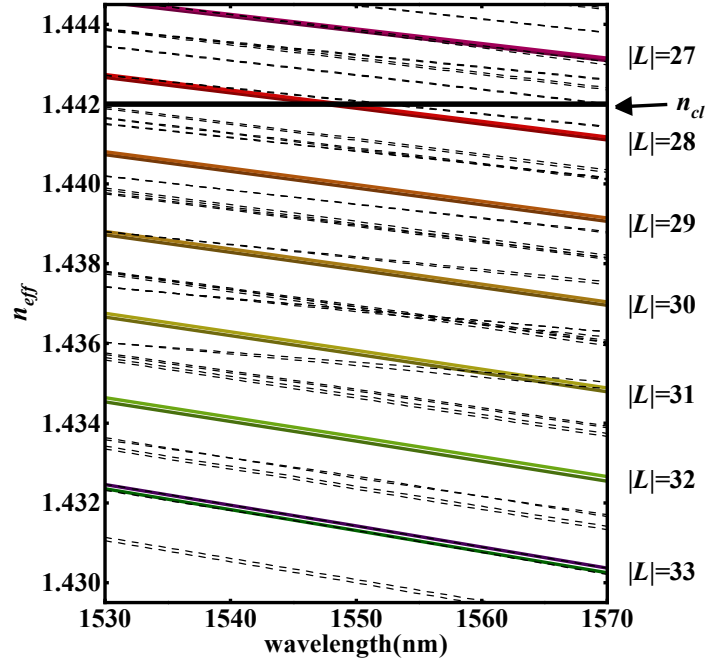

Fig. 8. $n_{eff}$ distribution curves of OAM modes versus wavelength in the low-loss SI-RCF. The solid black line represents the refractive index of the ring-core fiber cladding, the colored solid lines represent the first-order radial modes with different topological charges, and the dashed black line indicates the undesired higher-order radial modes.

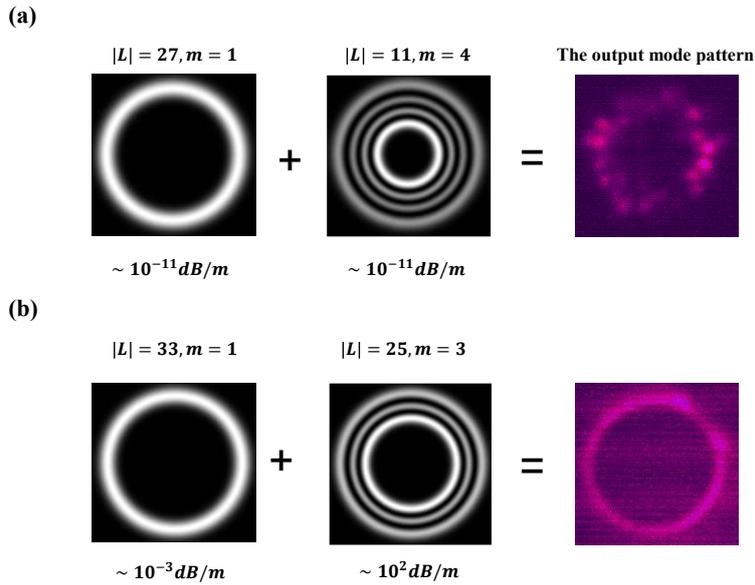

Fig. 9. Mechanism of topological confinement suppressing mode coupling in the low-loss SI-RCF. (a) Distorted coupled pattern of the $|L|=27, m=1$ mode and the $|L|=11, m=4$ mode. (b) Clear annular pattern of the $|L|=33, m=1$ mode and the $|L|=25, m=3$ mode.

The experimentally observed mode patterns of TCMs with different topological charges after long-distance transmission in the low-loss SI-RCF are shown in Fig. 4.6. It is verified that TCMs maintain clear annular patterns even after long-distance transmission of 3 km. The characteristics of the TCMs supported by the low-loss SI-RCF are listed in Table 1. The fiber supports 24 low-loss TCMs with $|L|$ from 28 to 33. The transmission loss was measured using the cut-back method. The results indicate that the transmission loss of these TCMs is no more than 2.6 dB/km, with the lowest loss of approximately 0.41 dB/km achieved for $|L|=28$ mode. In addition, $\Delta n_{eff}$ between the SOa and SOaa modes are all greater than $5\times10^{-5}$ and the $\Delta n_{eff}$ between adjacent first-order radial modes reach the order of $10^{-3}$. Such large $\Delta n_{eff}$ can fully satisfy the requirements for efficient separation of degenerate modes.

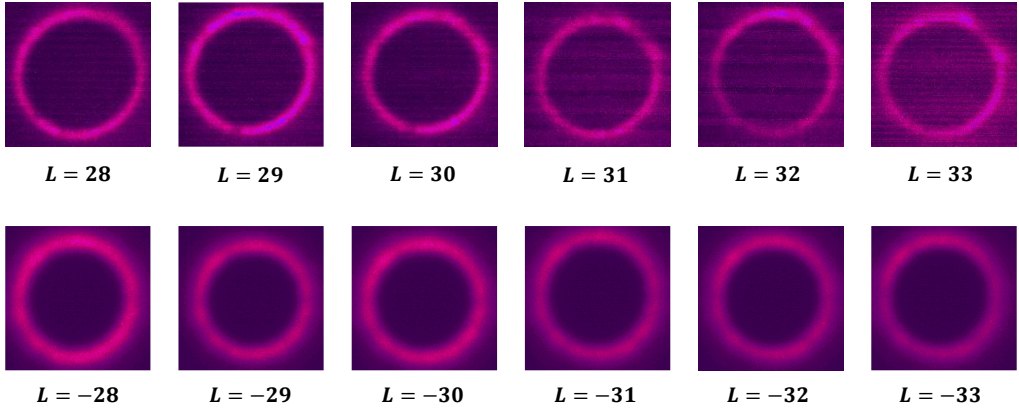

Fig. 10. Mode patterns of TCMs with different topological charges after 3km transmission.

Table 1. The characteristics of the TCMs supported by the low-loss SI-RCF

| TCMs | Loss(dB/km) | $\Delta n_{eff}$ between SOa modes and SOaa modes ($\times 10^{-5}$) | $\Delta n_{eff}$ between adjacent first-order radial modes ($\times 10^{-3}$) |
|---|---|---|---|
| $|L|=28$ | ~0.41 | 6.85 | 1.93 |
| $|L|=29$ | ~0.49 | 7.58 | 1.99 |
| $|L|=30$ | ~0.64 | 8.37 | 2.06 |
| $|L|=31$ | ~0.64 | 9.21 | 2.12 |
| $|L|=32$ | ~0.67 | 10.10 | 2.18 |
| $|L|=33$ | ~2.53 | 6.85 | 2.24 |

## 4. Maximum SNR and spectral efficiency estimation

The SNR of the fiber-optic communication system is affected by fiber nonlinearity, which can be approximated using the Gaussian model of nonlinear interference (NLI) power [26]. In addition, amplified spontaneous emission (ASE) noise represents another key factor limiting the SNR. Therefore, the expression for the maximum SNR in a crosstalk-free fiber can be written as [27]:

$$SNR_{SC,max,dB} \approx \frac{1}{3}\left[10log_{10}(|\beta_2|L_{eff}) - 20log_{10}(\gamma L_{eff}) - 2\alpha_{dB}L_s\right] + 10log_{10}\left(\frac{C_{system}}{N_s}\right) \quad (3)$$

where $\beta_2 = -\lambda^2 D/(2\pi c)$ is related to chromatic dispersion $D$, $L_{eff} = [1-\exp(-\alpha L_s)]/\alpha$ is the effective length, $\gamma = 2\pi n_2/(\lambda A_{eff})$ is the nonlinear coefficient, $n_2$ is the nonlinear refractive index determined by the material, $\alpha$ is the power loss per unit length, $L_s$ is the span length. The system term

is given by

$$\frac{C_{system}}{N_s} = \left[\ln\left(\pi^2 |\beta_2| L_{eff} B_{WDM}^2\right)\right]^{\frac{1}{3}} \left\{\left[\left(\frac{2}{\pi}\right)^{\frac{1}{3}} N_s (Fh\nu)^{\frac{2}{3}}\right]\right\}^{-1} \quad (4)$$

where $B_{WDM} = N_{ch} R_s$ is the WDM bandwidth, $N_{ch}$ is number of WDM channels, $R_s$ is the symbol rate, $N_s$ is the number of spans, $F$ is the EDFA noise figure, $h$ is the Planck constant, and $\nu$ is the center frequency of WDM comb. The three main factors affecting the maximum SNR of the system are $A_{eff}$, $D$, and $\alpha$. For single-mode fiber (SMF), these parameters are 80 μm², 16.8 ps/nm/km and 0.19 dB/km, respectively. Figure 11 shows the $A_{eff}$ and $D$ of TCMs in the low-loss SI-RCF over the C+L band, with five selected wavelengths: 1530 nm, 1550 nm, 1565 nm, 1590 nm, and 1625 nm.

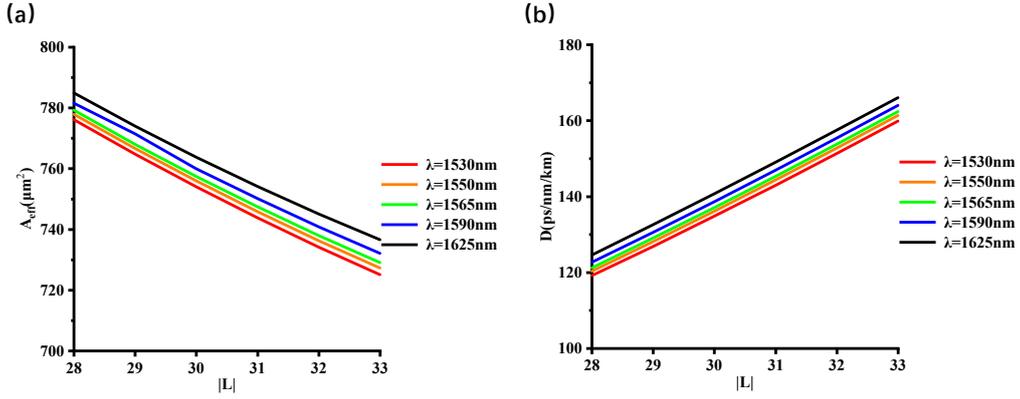

Fig. 11. $A_{eff}$ and $D$ of TCMs in the low-loss SI-RCF over the C+L band.

Figure 12 shows the maximum SNR versus transmission distance at 1550 nm for the fundamental mode in SMF and the TCMs in the low-loss SI-RCF. Although the transmission loss of the TCMs in the fiber is still higher than that of the fundamental mode in SMF, which limits its application in longer distance transmission. However, the TCMs exhibit lower nonlinear effects, leading to a maximum SNR advantage of approximately 9 dB over short distances. This SNR advantage can be maintained up to about 100 km for TCMs with $|L| = 28 - 32$. For the mode with $|L| = 33$, its transmission loss is around 2.53 dB/km, causing a rapid decay of the maximum SNR over short distances, so its SNR advantage only persists up to about 10 km.

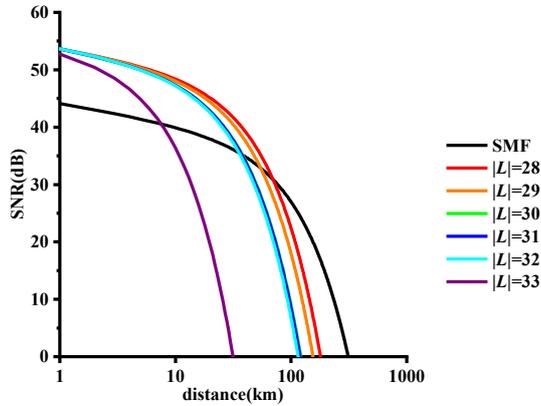

Fig. 12. The maximum SNR versus transmission distance at 1550 nm for the fundamental mode in SMF and the TCMs in the low-loss SI-RCF.

We calculate the spectral efficiency of individual TCMs and the total spectral efficiency of the low-loss SI-RCF. Notably, this estimation is based on ideal conditions, assuming sufficiently advanced device technologies to minimize crosstalk, no nonlinear mode interactions within the fiber, and full utilization of each spatial channel by multiplexers and demultiplexers. Figure 13 shows the spectral efficiency versus transmission distance at 1550 nm for the TCMs in the low-loss SI-RCF. It is concluded that the spectral efficiency of all topologically confined modes reaches 40 bit/s/Hz over a 10 km fiber span. For a 100 km span, the spectral efficiency of all TCMs reaches above 10 bit/s/Hz except for the mode with $|L|=33$. The total spectral efficiency of the ring-core fiber reaches 364bit/s/Hz, and reaches at 89 bit/s/Hz over a 100 km distance. In addition, the total spectral efficiency of a 22-core multi-core fiber is also plotted for comparison. The spectral efficiency of this multi-core fiber is approximately 211 bit/s/Hz at 30km [28], whereas the theoretical total spectral efficiency of the employed ring-core fiber at the same distance is about 269 bit/s/Hz, demonstrating a higher spectral efficiency than the 22-core MCF.

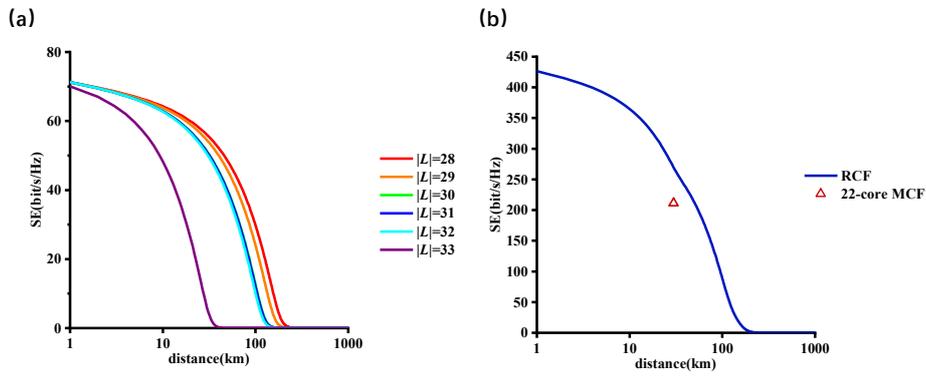

Fig. 13. The spectral efficiency of individual TCMs and the total spectral efficiency of the low-loss SI-RCF.

## 5. Conclusion

In this paper, we investigate the effects of outer ring radius $r_2$, refractive index difference $\Delta n$ and wavelength $\lambda$ on the characteristics of TCMs. The minimum thresholds of $r_2$ and $\Delta n$ required to realize the topological confinement mechanism at a wavelength of 1550 nm are determined. And the transmission performance of TCMs in GI-RCFs is analyzed, which exhibits superior nonlinear performance compared with SI-RCFs. We fabricate a type pf low-loss SI-RCF supporting the topological confinement mechanism. The overall transmission loss of the TCMs is no more than 2.6 dB/km, with a minimum value of only 0.41 dB/km. The $\Delta n_{eff}$ between the SOa and SOaa modes exceeds $5\times10^{-5}$, and $\Delta n_{eff}$ between adjacent first-order radial modes is on the order of $10^{-3}$. According to model estimation, the maximum SNR of all TCMs after short-distance transmission is improved by approximately 9 dB compared with the fundamental mode in standard SMF. The ideal SE reaches 364 bit/s/Hz, demonstrating a significant advantage in transmission capacity.